\documentclass[MSNbibl,number,citesort,dvips]{arxstspdf}
\usepackage{flushend}
\usepackage{stfloats}
\volume{28}
\issue{2}
\pubyear{2013}
\firstpage{257}
\lastpage{268}
\doi{10.1214/13-STS415} 

\makeatletter

\newtheorem{theorem}{Theorem}

\newproclaim{definition}{Definition}

\newcommand{\btheta}{{\bolds\theta}}
\newcommand{\bphi}{{\bolds\phi}}

\newcommand{\bYobs}{\mathbf Y_{\mathrm{obs}}}
\newcommand{\bYmis}{\mathbf Y_{\mathrm{mis}}}
\newcommand{\bM}{\mathbf M}
\newcommand{\bX}{\mathbf X}
\newcommand{\bY}{\mathbf Y}

\newcommand{\bem}{\mathbf m}
\newcommand{\bu}{\mathbf u}
\newcommand{\byactual}{\tilde{\mathbf y}}
\newcommand{\bmactual}{\tilde{\mathbf m}}
\newcommand{\mis}{\bar{o}}

\makeatother

\begin{document}
\begin{frontmatter}

\title{What Is Meant by ``Missing at Random''?}
\runtitle{Missing at Random}

\begin{aug}
\author[a]{\fnms{Shaun} \snm{Seaman}\corref{}\ead[label=e1]{shaun.seaman@mrc-bsu.cam.ac.uk}},
\author[b]{\fnms{John} \snm{Galati}},
\author[a]{\fnms{Dan} \snm{Jackson}}
\and
\author[c]{\fnms{John} \snm{Carlin}}
\runauthor{Seaman, Galati, Jackson and Carlin}

\affiliation{MRC Biostatistics Unit, Clinical Epidemiology and Biostatistics Unit
and La Trobe University, MRC Biostatistics Unit, and Clinical
Epidemiology and Biostatistics Unit and University
of Melbourne}

\address[a]{Shaun Seaman is Senior Statistician and Dan Jackson is Senior Statistician,
MRC Biostatistics Unit, Cambridge, United Kingdom \printead{e1}.}
\address[b]{John Galati is Senior Research Officer, Clinical Epidemiology and
Biostatistics Unit, Murdoch Childrens Research Institute, and
Department of Mathematics and Statistics, La Trobe University,
Victoria, Australia.}
\address[c]{John Carlin is Director, Clinical Epidemiology and
Biostatistics Unit, Murdoch Childrens Research Institute and University
of Melbourne, Victoria, Australia.}

\end{aug}

%
\begin{abstract}
The concept of missing at random is central in the literature on
statistical analysis with missing data. In general, inference using
incomplete data should be based not only on observed data values but
should also take account of the pattern of missing values. However, it
is often said that if data are missing at random, valid inference using
likelihood approaches (including Bayesian) can be obtained ignoring the
missingness mechanism. Unfortunately, the term ``missing at random'' has
been used inconsistently and not always clearly; there has also been a
lack of clarity around the meaning of ``valid inference using
likelihood''. These issues have created potential for confusion about
the exact conditions under which the missingness mechanism can be
ignored, and perhaps fed confusion around the meaning of ``analysis
ignoring the missingness mechanism''. Here we provide standardised
precise definitions of ``missing at random'' and ``missing completely at
random'', in order to promote unification of the theory. Using these
definitions we clarify the conditions that suffice for ``valid
inference'' to be obtained under a variety of inferential paradigms.
\end{abstract}

%
\vspace*{-6pt}

\end{frontmatter}

\section{Introduction}

The literature on missing data is not entirely clear with respect to
the assumptions required for different types of analysis to be valid.
First, although the term ``missing at random'' (MAR) has been widely
regarded as central to the theory underlying missing data methods since
the seminal paper of Rubin (1976)~\cite{Rubin1976}, it has not always been used in a
consistent manner.
There has often been a lack of detail about whether the MAR condition
is a statement only about the realised missingness pattern or about all
possible patterns and whether it is only about the realised values of
the observed data or all possible observable data values.
Second, the distinction between direct-likelihood and frequentist
inference using the likelihood function is not always made clear.
Third, it is sometimes said that ``missing completely at random'' (MCAR)
is needed for frequentist inference; at other times MAR is said to be
sufficient.

While it is clear that some researchers writing on the theory of
missing data have known what they intended, the omission of details by
many authors, together with the seemingly different conditions assumed
by different authors, make it difficult for readers to know precisely
what was meant, and also to compare the work of different authors. This
confusion has implications for statistical practice, since data
analysts are encouraged to consider the plausibility of the MAR
assumption before applying certain methods of analysis (e.g.,
\cite{Sterne2009}), but if the conscientious analyst consults the
theoretical literature they will struggle to find a clear consensus on
definitions and on how they relate to the validity of possible analytic
approaches. Further confusion surrounds the concept of
``ignorability'', which does not seem to be well understood by
practitioners and may be misinterpreted as providing a broad licence to
ignore the fact that not all the desired data have been observed.

In the present article, our objectives are to:\break (1)~draw attention to
the various gaps and inconsistencies in some definitions of MAR used in
the literature; (2)~provide unambiguous formulations of relevant MAR
definitions; and (3) explain the relation between MAR and ignorability
under different frameworks of statistical inference and, in so doing,
identify the need for more than one definition of MAR.

The structure of the paper is as follows.
In Section~\ref{sectmydefinitions} we provide definitions of two
distinct MAR conditions, one stronger than the other, and likewise for MCAR.
The inconsistency in previous usage of the terms ``MAR'' and ``MCAR'' is
documented in Section~\ref{sectreview}.
The definitions of MAR and MCAR are central to the concept of
ignorability, the definition of which varies according to the chosen
framework of statistical inference.
In Section~\ref{sectinferencetypes} we distinguish between
direct-likelihood inference, Bayesian inference, frequentist inference
using the likelihood function and the frequentist properties of
Bayesian estimators.
Section~\ref{sectignorability} contains an explanation of which
MAR/MCAR conditions are needed for the missingness mechanism to be
ignorable for each of these types of inference.
Section~\ref{sectconditional} covers the use of conditional
likelihood and repeated sampling.
We end with a discussion.

\section{Two Definitions of MAR and MCAR}
\label{sectmydefinitions}

We use $\bY$ to denote the vector of potentially observable data
values (on all sample units), which for modelling purposes we treat as
a random variable.
Let $\bM$ denote a vector of missingness indicators of the same length
as $\bY$.
The $j$th element of $\bM$ equals one if the $j$th element of $\bY$
is observed and zero if it is missing.
Let $o(\bY, \bM)$, a function of $\bY$ and $\bM$, denote the
subvector of $\bY$ consisting of elements whose corresponding elements
of $\bM$ equal one.
So, $o(\bY, \bM)$ contains the observed elements of $\bY$.
Let $K$ denote the length of $o(\bY, \bM)$.
So, $K$ is a random variable and is equal to the sum of the
elements of $\bM$.
When no elements of $\bY$ are observed, $o(\bY, \bM)$ is the
empty set and $K=0$.
The reader may be familiar with the notation $\bYobs$ and $\bYmis$.
We choose not to use this notation because it is ambiguous, as we
explain in Section~\ref{sectreview}.
However, our notation $o(\bY, \bM)$ is equivalent to $\bYobs$ as
usually interpreted.
When we consider a specific sample, it is convenient to have notation
for the realised values of the random variables $\bM$ and $\bY$; we
denote these realised values as $\bmactual$ and $\byactual$, respectively.
``Realised'' and ``observed'' values should not be confused.
The observed value, $o(\bY, \bM)$, of $\bY$ is a random variable and
has a realised value, $o(\byactual, \bmactual)$.
The values of $\bmactual$ and $o(\byactual, \bmactual)$ are
known, but that of $\byactual$ is only known if all elements of
$\bmactual$ equal one.

In the special case where the data are modelled as a set of $J$ random
variables measured on each of $n$ units, as is often the case, $\bY$
is a vector of length $nJ$.
Although in this special case one might alternatively define $\bY$ as
a matrix with $n$ rows and $J$ columns, for the sake of generality we
do not do this.
For example, suppose that $\bY$ consists of two random variables, $X$
and $Z$, measured on each of two units, that the realised value of $(X,
Z)$ is $(10,3)$ for the first unit and $(4,2)$ for the second, and that
$X$ is observed for both units but $Z$ is only observed for the second.
Then $\byactual= (10,3,4,2)^T$, $\bmactual= (1,0,1,1)^T$ and $o
(\byactual, \bmactual) = (10,4,2)^T$.
Note that $o(\mathbf y, \bem)$ cannot be interpreted without the
accompanying value of~$\tilde{\bem}$.

Consider a hypothesised ``missingness model'', that is, a model for the
conditional distribution of $\bM$ given~$\bY$.
Let $g_{\phi} (\bem\mid\mathbf y )$ denote the probability that $\bM=
\bem$ given that $\bY= \mathbf y $ according to this model, where $\bphi
$ is an unknown parameter.
We now present two definitions of MAR.

\begin{definition}\label{defin1}
The data are realised MAR if~$\forall\bphi$,
\begin{eqnarray}
&&
g_{\phi} (\bmactual\mid\mathbf y ) = g_{\phi} (\bmactual\mid
\byactual)\nonumber\\
&&\eqntext{\forall\mathbf y \mbox{ such that }o(\mathbf y, \bmactual) =
o(\byactual, \bmactual)}
\end{eqnarray}
(where $\mathbf y $ represents a value of $\bY$).
This means that the hypothesised missingness model always (i.e., for
all values of $\bphi$) assumes that the conditional probability that
the missingness pattern $\bM$ is its realised value $\bmactual$,
given the realised values of the elements of the data $\bY$ that are
observed when $\bM= \bmactual$ and the values of the remaining,
missing, elements, does not depend on these missing elements.
Rubin~\cite{Rubin1976}\vadjust{\goodbreak} expressed this as follows: ``The missing
data are missing at random if for each possible value of the parameter
$\bphi$, the conditional probability of the observed pattern of
missing data, given the missing data and the value of the observed
data, is the same for all possible values of the missing data''.
There are several things to note about this definition.
First, it is a statement only about the realised missingness pattern
and realised observed data, not about missingness patterns or observed
data that could have been realised but were not.
Second, it is a statement about a hypothesised missingness model,
rather than necessarily the true missingness process.
\end{definition}

\begin{definition}\label{defin2}
The data are everywhere MAR if~$\forall\bphi$,
\begin{eqnarray}
&&g_{\phi} (\bem\mid\mathbf y ) = g_{\phi} \bigl(\bem\mid\mathbf y ^*\bigr)\qquad\qquad\nonumber\\
&&\eqntext{\forall\bem, \mathbf y, \mathbf y ^*\mbox{ such that }o(\mathbf y, \bem) =
o\bigl(\mathbf y ^*, \bem\bigr)}
\end{eqnarray}
(where $\mathbf y $ and $\mathbf y ^*$ represent a pair of values of $\bY$).
This means that the hypothesised missingness model always assumes that,
for any value of the data, the probability of any possible missingness
pattern, given the values of the corresponding observed elements and
missing elements of the data, does not depend on the values of the
missing elements.
In order to make more obvious the difference between realised and
everywhere MAR, note that Definition~\ref{defin1} can be rewritten as follows.
The data are realised MAR if $\forall\bphi$, $g_{\phi} (\bmactual
\mid\mathbf y ) = g_{\phi} (\bmactual\mid\mathbf y ^*)$ $\forall\mathbf y,
\mathbf y ^*$ such that $o(\mathbf y,\break \bmactual) = o(\mathbf y
^*,
\bmactual) = o(\byactual, \bmactual)$.
Unlike realised MAR, everywhere MAR is a statement about all possible
missingness patterns and values of the observed data.
Note that everywhere MAR implies realised MAR.
\end{definition}

To illustrate and clarify the notation that we have used here, consider
the example given above, that is, $\byactual= (10,3,4,2)^T$,
$\bmactual= (1,0,1,1)^T$ and $o(\byactual, \bmactual) = (10,4,2)^T$.
The data are realised MAR if $\forall\bphi$, $g_{\phi} ( (1,0,\allowbreak1,1)^T
\mid\mathbf y ) = g_{\phi} ( (1,0,1,1)^T \mid\mathbf y ^*)$ $\forall\mathbf y,
\mathbf y ^*$ such that the first, second and fourth elements of both $\mathbf
y $ and $\mathbf y ^{*}$ equal, respectively, 10, 4 and 2.
That is, the data are realised MAR if, for any $\bphi$, $g_{\phi} (
(1,0,1,1)^T \mid(10, a,\break 4, 2)^T ) = g_{\phi} ( (1,0,1,1)^T \mid(10,
b, 4, 2)^T )$ for all $a, b$ in the sample space of the second element
of~$\bY$.

Now consider the special case of independent identically distributed
(i.i.d.) data, that is, $\bY= (\bY_1^T,\ldots,\allowbreak \bY_n^T)^T$ and
$\bM\,{=}\, (\bM_1^T,\ldots, \bM_n^T)^T$, where $(\bY_i, \bM_i)$
($i\,{=}\break1,\ldots, n)$ are i.i.d.
Let $o_1(\bY_i, \bM_i)$ denote the subvector of $\bY_i$
consisting of elements of $\bY_i$ whose corresponding elements of $\bM
_i$ equal one.
[Note that the function $o_1$\vadjust{\goodbreak} is analogous to the previously defined
$o(\cdot)$, but whereas $o(\cdot)$ is a function of all the data, $o_1$ is a
function of only the data for a single unit.]
So, $\bY_i$, $\bM_i$ and $o_1(\bY_i, \bM_i)$ denote the data,
the missingness pattern and the observed data, respectively, for the
$i$th of $n$ units.
Consider a hypothesised model for the conditional distribution of $\bM
_i$ given $\bY_i$, and let $g_{\phi, 1} (\bem_i \mid\mathbf y _i)$
denote the probability that $\bM_i = \bem_i$ given that $\bY_i = \mathbf
y _i$ according to the model.
In this case, Definitions~\ref{defin2} and~\ref{defin3} are equivalent.

\begin{definition}\label{defin3}
The data are everywhere MAR if $\forall i, \bphi$,
\begin{eqnarray}
g_{\phi, 1} (\bem_i \mid\mathbf y _i) = g_{\phi, 1} \bigl(\bem_i \mid\mathbf
y _i^*\bigr)\qquad\qquad\nonumber\\
&&\eqntext{\forall\mathbf y _i, \mathbf y _i^*\mbox{ such that }o_1(\mathbf y _i,
\bem_i) = o_1\bigl(\mathbf y _i^*, \bem_i\bigr).}
\end{eqnarray}
\end{definition}

Definition~\ref{defin3} may only be applied when $(\bY_1, \bM_1),\allowbreak\ldots, (\bY
_n, \bM_n)$ are i.i.d.
If, for example, $\bY_1,\ldots, \bY_n$ were i.i.d. and $\bM_1,\ldots, \bM_n$ were identically
distribut\-ed but with $\bM_i$
depending on $\bM_j$ and/or $\bY_j$ for $j \neq i$, then $(\bY_1,
\bM_1),\ldots, (\bY_n, \bM_n)$ would not be i.i.d. and so
Definition~\ref{defin3} could not apply.
The data might nevertheless still be everywhere MAR by Definition~\ref{defin2}.

Finally, we present two definitions of MCAR.

\begin{definition}\label{defin4}
The data are realised MCAR\break if~$\forall\bphi$,
\[
g_{\phi} (\bmactual\mid\mathbf y ) = g_{\phi} \bigl(\bmactual\mid\mathbf y
^*\bigr)\quad\forall\mathbf y, \mathbf y ^*.
\]
\end{definition}

\begin{definition}\label{defin5}
The data are everywhere MCAR if~$\forall\bphi$,
\[
g_{\phi} (\bem\mid\mathbf y ) = g_{\phi} \bigl(\bem\mid\mathbf y ^*\bigr)
\quad\forall\bem, \mathbf y, \mathbf y ^*.
\]
\end{definition}

Realised MCAR means that the probability of the realised missingness
pattern given the data does not depend on the data.
Realised MCAR implies realised MAR but not everywhere MAR.
Everywhere MCAR means that the probability of \textit{any} missingness pattern
given the data does not depend on the data, that is, $\bM$ is
independent of $\bY$.
Everywhere MCAR implies realised MCAR, realised MAR and everywhere MAR.

\section{MAR and MCAR in the Literature: A~Review}
\label{sectreview}

Historically, the first definition of MAR was that of Rubin
(1976)~\cite{Rubin1976}.
This is Definition~\ref{defin1}, that is, the definition for realised MAR (apart
from minor differences in notation and the fact that Rubin's definition
begins ``The missing data are MAR'' rather than ``The data are MAR'').
Rubin (1987)~\cite{Rubin1987} largely avoided\vadjust{\goodbreak} the term ``MAR'',
preferring instead the\break terms ``ignorable sampling'' and ``ignorable response''.
However, he did (page 53) briefly discuss the relation between these
three terms.
It is evident from that discussion that he was using the Rubin (1976)~\cite{Rubin1976}
definition.
Heitjan and colleagues, in a series of papers (e.g., \cite
{Heitjan1991,Heitjan1993,Heitjan1994,Heitjanletter1994,Heitjan1996,Heitjan1997}),
consistently used ``MAR'' to\break mean realised MAR.
Harel and Schafer~\cite{Harel2009} also defined realised MAR.
Most other authors have used ``MAR'' to mean everywhere MAR.

Several authors (Schafer~\cite{Schafer1997}; Kenward and
Molenberghs~\cite{Kenward1998}; Lu and Copas~\cite{Lu2004};
Jaeger~\cite{Jaeger2005}) provided definitions of everywhere MAR
but accompanied this definition with a citation of Rubin (1976) \cite{Rubin1976} (which
defines realised, rather than everywhere, MAR).
In fact, most of these authors said explicitly that their definition
was an expression of Rubin's (1976)~\cite{Rubin1976} definition.

The potential of the variety of definitions of MAR to cause confusion
was illustrated by an exchange of letters between Heitjan~\cite{Heitjanletter1994}
and Diggle~\cite{Diggleletter1994}.
Note that according to Rubin's (1976)~\cite{Rubin1976} definition (i.e., realised MAR),
if all the data are observed, they cannot fail to be MAR (although one
might alternatively say that his definition is a statement about
\textit{the missing data} and in this situation there are no missing data, so
there are no missing data to be MAR).
Heitjan gave an example in which a single variable $X$ is measured on
$n$ individuals and could potentially be missing on some of these individuals.
However, he supposed that in the data set actually observed, $X$ is
observed on all $n$ individuals, so there are no missing data.
He stated that the data are MAR.
Diggle responded by saying that the data are not MAR, since the
probability that $X$ is observed depends on $X$, which could be missing.
The reason for this disagreement is that Heitjan was using the
definition of realised MAR whereas Diggle was using that of everywhere MAR.

In addition to the problems caused by this dual use of the term ``MAR'',
definitions of MAR found in some of the key literature on missing data,
including textbooks, contain certain ambiguities.

Many authors (Little and Rubin~\cite{Little1987,Little2002}; Scha\-fer~\cite{Schafer1997}; Kenward and Molenberghs~\cite{Kenward1998}; Harel
and Scha\-fer~\cite{Harel2009}; Fitzmaurice et al.
\cite{Fitzmaurice2011}) used the problematic notation $\bYobs$ and
$\bYmis$ mentioned in Section~\ref{sectmydefinitions}. Little and Rubin
\cite{Little1987,Little2002}, for example, said that $\bYobs$ denotes
the observed components or entries of $\bY$, that $\bYmis$ denotes the
missing components, and that the missing data mechanism is called MAR
if
%
\begin{equation}\label{eqMARRubin2002}
f(\bM\mid\bY, \bphi) = f(\bM\mid\bYobs, \bphi) \quad\forall\bYmis, \bphi
\end{equation}
[where $f(\cdot \mid\cdot)$ denotes a conditional distribution].
The notation $f(\bM\mid\bYobs, \bphi)$ is somewhat confusing,
because $\bYobs$ is itself a function of $\bM$.
Interpreted literally, $\bYobs= o(\bY, \bM)$.
Hence, if $\bYobs$ is known, then $K$ is also known, and so
$f(\bM\mid\bYobs, \bphi)$ should equal zero unless the number of
nonzero elements of $\bM$ equals~$K$.
Nevertheless, we presume that equation (\ref{eqMARRubin2002}) was
intended to mean Definition~\ref{defin2} (i.e., everywhere MAR).
Fitzmaurice et al.~\cite{Fitzmaurice2011} gave a definition
similar to equation (\ref{eqMARRubin2002}), but added that this
means $\bM$ is conditionally independent of $\bYmis$ given $\bYobs$.
This is rather difficult to interpret, given that $\bYmis$ is a
function of $\bM$.

Another source of ambiguity concerns the parameter $\bphi$.
Definitions~\ref{defin1}--\ref{defin3} require a particular equality to hold for all values
of $\bphi$.
Several authors (Robins and Gill~\cite{Robins1997}; Kenward and
Molenberghs~\cite{Kenward1998}; Tsiatis~\cite{Tsiatis2006};
Fitzmaurice et al.~\cite{Fitzmaurice2011}) omitted the parameter
$\bphi$ when defining MAR, with the result that it is not obvious
whether equality is required to hold for all $\bphi$ or just for its
``true'' value.
Schafer~\cite{Schafer1997} did include $\bphi$, but was also
unclear about whether equality must hold for all $\bphi$.
Judging from the use that these authors made of their MAR assumptions,
most of them seem implicitly to have meant that the equality should
hold for all $\bphi$.
However, Fitzmaurice et al.~\cite{Fitzmaurice2011} seem to
require equation (\ref{eqMARRubin2002}) to hold only for the true
value of $\bphi$: they appear to be referring to the ``true''
missingness mechanism, rather than to a model for the missingness.
We shall return to this point in Section~\ref{sectdiscussion}.

Just as there can be ambiguity about $\bphi$, it is sometimes not
entirely clear whether a definition of MAR requires an equality to hold
for all $\bY$ or just for $\bY$ compatible with $o(\byactual, M)$.
See, in particular, equation (\ref{eqMARRubin2002}).

We have concentrated on MAR, but there is also ambiguity about the
definition of MCAR. In his original 1976 paper~\cite{Rubin1976}, Rubin
did not mention\break MCAR. He instead introduced the concept of the observed
data being ``observed at random''. The realised\break MCAR definition
(Definition~\ref{defin4}) is equivalent to the combination of the
missing data being realised MAR and the observed data being observed at
random~\cite{Heitjan1994} (see also Little~\cite{Little1976}). Heitjan
and colleagues have used ``MCAR'' to mean realised MCAR. Many other
authors (e.g., Little and Rubin~\cite{Little1987,Little2002} and
\cite{Schafer1997}) have used ``MCAR'' to mean everywhere MCAR.
In the situation of repeated-measures outcome data with fully observed\vadjust{\goodbreak}
covariates, Molenberghs and Kenward~\cite{MolenberghsKenward2007} used
``MCAR'' to mean that missingness in the outcomes cannot depend on the
outcomes but can depend on the covariates. Elsewhere this has been
called ``covariate-dependent MCAR''~\cite{Little1995,Wood2004}.

\section{Direct-Likelihood, Bayesian and Frequentist Inference}
\label{sectinferencetypes}

In Section~\ref{sectignorability} we shall discuss ignorability.
The definition of ignorability depends on the framework of inference adopted.
Here we review the distinctions between four types of inference:
Bayesian inference, direct-likelihood inference (also known as
pure-likelihood inference), general frequentist inference and
frequentist likelihood inference.
For simplicity of exposition, we describe inference when the data $\bY
$ are fully observed.
In Section~\ref{sectignorability} we describe the generalisation to
incomplete data.

In Bayesian and direct-likelihood inference a probability distribution
function is specified for the\break data~$\bY$.
This function involves a finite set of unknown parameters, $\btheta$.
Some of these are of interest and the aim is to make inference about
their values; others may be nuisance parameters.
The likelihood is defined as any multiple of this probability
distribution function where the multiplier does not depend on any of
the parameters.
Whereas the probability distribution function is regarded as a function
of the data with the values of the parameters considered fixed, the
likelihood is regarded as a function of the parameters with the data
considered fixed.

In direct-likelihood inference~\cite{Clayton1993,Pawitan2001,Reid2000},
the value of the parameters at which the likelihood is a maximum (the
maximum likelihood estimate) is used as a point estimate and the ratio
of the value of the likelihood at different parameter values is used to
judge which parameter values are plausible. The normalised likelihood
is defined as the likelihood divided by the value of the likelihood at
the maximum likelihood estimate (so that the normalised likelihood
takes value one at the maximum likelihood estimate). When there is only
one parameter, a likelihood interval is defined as the set of parameter
values within which the values of the normalised likelihood are greater
than some threshold. Different thresholds have been proposed, for
example, Fisher~\cite{Fisher1956} suggested $1/15$ and Royall
\cite{Royall1997} suggested $1/32$.

When there is more than one parameter, a likelihood interval for any
one of them can be obtained by first eliminating the others.
Two commonly used ways to eliminate parameters are the profile
likelihood method and the conditional likelihood method.
Suppose, without loss of generality, that $\btheta= (\theta_1,
\btheta_2)$, where $\btheta_2$ are the parameters to be eliminated.
The profile likelihood for $\theta_1$ is defined as the function
obtained, for each possible value of $\theta_1$, by fixing $\theta_1$
at that value and then maximising the likelihood for $\btheta$ over
the space of $\btheta_2$.
In the profile likelihood method, a likelihood interval for $\theta_1$
is calculated using the profile likelihood for $\theta_1$ in place
of the likelihood for $\btheta$.
In the conditional likelihood method, a~conditional probability
distribution function is specified for $\bY$ given a (possibly vector)
function of $\bY$.
The resulting conditional likelihood contains fewer parameters than
the unconditional likelihood, that is, that based on the unconditional
probability distribution function for $\bY$.
If the conditional likelihood contains only $\theta_1$, it can be
used to construct a likelihood interval for $\theta_1$.
If it contains additional parameters, these can be eliminated using the
profile likelihood method.
There is no clear theoretical basis for choosing between the profile
likelihood and conditional likelihood approaches, and each appear to
have their merits for different situations.

In Bayesian inference, uncertainty about parameters is represented
directly by probability models, requiring a prior distribution to be specified.
The posterior distribution of the parameters is obtained by Bayes' theorem.
For any of the parameters in the model, the mean of its posterior
distribution is typically used as a point estimate and $(\alpha_l,
\alpha_u)$ used as an interval of uncertainty (a credible interval),
where $\alpha_l$ and $\alpha_u$ are the $l$th and $u$th centiles
(e.g., 2.5th and 97.5th) of that parameter's marginal posterior distribution.
This interval is interpretable as meaning that the posterior
probability that the parameter lies within $(\alpha_l, \alpha_u)$ is
$(u-l) / 100$.
The use of the marginal posterior distribution means that all other
parameters are eliminated by integrating them out of the joint
posterior distribution of all the parameters.

In direct-likelihood inference and Bayesian inference as described
above, only the realised value of $\bY$ is of interest; there is no
consideration of other values of $\bY$ that could have been realised
but which were not. Frequentist inference, on the other hand, is
concerned with the (hypothetical) repeated sampling of $\bY$ and with
the properties of inferential summaries such as point and interval
estimates under this repeated sampling. It is only when repeated
sampling is considered that the concepts of bias, standard error,\vadjust{\goodbreak}
efficiency, power and confidence interval become meaningful. The bias
of an estimator of a parameter, for example, is defined as the
difference between the mean of the sampling distribution of the
estimator and the true value of the parameter; the standard error is
the standard deviation of the sampling distribution of the estimator;
a~confidence interval is an interval obtained using a rule that has a
stated probability of producing an interval containing the true value
of the parameter in a repeated sample. One important example of a rule
for constructing confidence intervals is the rule used in
direct-likelihood inference to construct likelihood intervals, that is,
a likelihood interval becomes, in the framework of frequentist
inference, a confidence interval.

In frequentist inference, a function $s(\bY)$ of $\bY$ is chosen
and its realised value, $s(\byactual)$, is compared with the
sampling distribution of $s(\bY)$, that is, the distribution of
$s(\bY)$ in repeated samples.
This sampling distribution may be conditional on the realised value of
a (possibly vector) function of $\bY$.
We distinguish between general frequentist inference, where $s(\bY
)$ can be any function of $\bY$, and frequentist likelihood inference,
where $s(\bY)$ depends on $\bY$ only through the likelihood of
$\bY$.
Frequentist likelihood inference includes using the observed or
expected information to estimate the standard error of the maximum
likelihood estimator (MLE), using this MLE and standard error to
construct a confidence interval, using likelihood intervals as
confidence intervals, and using likelihood-ratio, Wald and score tests.
Frequentist likelihood inference is like direct-likelihood inference,
in that it also uses the MLE and likelihood intervals, but goes beyond
it, in that it involves claims about the behaviour of the MLE and
likelihood intervals in repeated samples.
Frequentist likelihood inference is often referred to simply as
``likelihood inference''.

Often even statisticians using Bayesian methods are interested in
frequentist properties of their estimators, for example, the bias of
the posterior mean or the coverage of a credible interval
\cite{Rubin1984,Kass1996}.

The distinction between direct-likelihood inference and frequentist
likelihood inference has not always been made clear in the literature.
For example, Heitjan and Rubin~\cite{Heitjan1991} and Harel and
Schafer~\cite{Harel2009} referred to direct-likelihood inference
simply as ``likelihood inference''.
Molenberghs et al.~\cite{Molenberghs2011} appear to use the
term ``direct-likelihood analysis'' when writing about repeated sample
properties of the likelihood.
Also, the potential interest in frequentist properties of Bayesian\vadjust{\goodbreak}
estimators has rarely been mentioned in the literature on missing data,
except in the context of multiple imputation.

\section{Ignorability of the Missingness Mechanism}
\label{sectignorability}

In this section we clarify which missingness assumption suffices for
the missingness mechanism to be ignorable for each of the types of
inferences described in Section~\ref{sectinferencetypes}.
Intuitively, ``ignorable'' means that inferences obtained from a
parametric model for the data alone are the same as inferences obtained
from a joint model for the data and missingness mechanism.
To serve as a workable definition, one needs to say what is meant by
``the same'', and in the literature authors have not always been explicit
on this point.
We endeavour to be clear, but defer specification of our definitions to
the relevant subsections below.

Consider a joint parametric model for the complete data $\bY$ and
missingness pattern $\bM$.
Let\break $f_{\theta} (\mathbf y ) g_{\phi} (\bem\mid\mathbf y )$ denote the
joint distribution of $\bY$ and $\bM$ according to this model, and
let $\Omega_{\theta, \phi}$ denote the joint parameter space for
$(\btheta, \bphi)$.
Let $\byactual$ and $\bmactual$ be a given realisation of $\bY$ and
$\bM$.
Let $\Omega_{\theta} = \pi_1( \Omega_{\theta, \phi} )$ and
$\Omega_{\phi} = \pi_2 ( \Omega_{\theta, \phi} )$ be the
parameter spaces for $\btheta$ and $\bphi$, respectively,
corresponding to the joint parameter space $\Omega_{\theta, \phi}$.
Following Heitjan and Basu~\cite{Heitjan1996}, we avoid
measure-theoretic difficulties by assuming that $\mathbf Y$ is discrete.
Because in reality all data are measured to finite precision, this
assumption is not restrictive.
Reference to continuous distributions should be interpreted as meaning
discrete distributions on a fine grid, and integrals can be interpreted
as sums.

The \textit{joint likelihood for $(\btheta, \bphi)$} is the function
with domain $\Omega_{\theta, \phi}$ given by
%
\begin{equation}\label{eqL1}
L_1 (\btheta, \bphi) = \int f_{\theta} (\mathbf y )
g_{\phi} (\bmactual\mid\mathbf y ) r(\mathbf y, \byactual, \bmactual) \,d \mathbf
y,
\end{equation}
%
where $r(\mathbf y, \byactual, \bmactual)$ equals one if $o(\mathbf y,
\bmactual) = o
(\byactual, \bmactual)$ and zero otherwise.
Note that the integral here integrates out the missing data.
The \textit{likelihood for $\btheta$ ignoring the missing-data mechanism}
is the function with domain $\Omega_{\theta}$ given by
%
\begin{equation}\label{eqL2}
L_2 (\btheta) = \int f_{\theta} (\mathbf y ) r(\mathbf y,
\byactual, \bmactual) \,d \mathbf y.
\end{equation}
For any fixed $\bphi\in\Omega_{\phi}$, the \textit{fixed-$\bphi$
likelihood for $\btheta$} is the function with domain $\Omega_{\theta
}$ given by
%
\begin{eqnarray}\label{eqL3}
L_{3, \phi} (\btheta) &=& \delta\bigl\{(\btheta,\bphi),\Omega_{\theta,\phi}
\bigr\}\nonumber\\[-8pt]\\[-8pt]
&&{}\cdot \int f_{\theta} (\mathbf y ) g_{\phi} (\bmactual\mid\mathbf y ) r(
\mathbf y, \byactual, \bmactual) \,d \mathbf y,\nonumber
\end{eqnarray}
where $\delta\{ (\btheta, \bphi), \Omega_{\theta, \phi} \}$
equals one if $(\btheta, \bphi) \in\Omega_{\theta, \phi}$ and
zero otherwise.
The \textit{profile likelihood for $\btheta$} is the function with domain
$\Omega_{\theta}$ given by
%
\begin{eqnarray}\label{eqL4}\quad
L_4 (\btheta) &=& %
\max_{\bphi\in\Omega_{\phi}} \biggl[
\delta\bigl\{(\btheta,\bphi),\Omega_{\theta,\phi}\bigr\} \nonumber\\[-8pt]\\[-8pt]
&&\hspace*{25.5pt}{}\cdot\int f_{\theta
} (
\mathbf y ) g_{\phi} (\bmactual\mid\mathbf y ) r(\mathbf y, \byactual,
\bmactual)
\,d \mathbf y \biggr].\nonumber
\end{eqnarray}
In Section~\ref{sectconditional} we shall consider the use of
conditional likelihoods.

\subsection{Direct-Likelihood Inference}
\label{sectignorabilitydirectlike}

The main work on ignorability for direct-likelihood inference can be
summed up in the following theorem.
After giving a proof, we shall discuss why this theorem has been
considered to justify the use of $L_2$, the likelihood for $\btheta$
ignoring the missing-data mechanism, when the data are realised MAR and
the parameters are distinct.

\begin{theorem}\label{theo1}
If realised MAR holds and $\Omega_{\theta, \phi} = \Omega_{\theta}
\times\Omega_{\phi}$, then: \textup{(i)} $L_1 (\btheta, \bphi)$ factorises
into two components, such that each parameter appears in only one
component; \textup{(ii)} for any $\bphi\in\Omega_{\phi}$ satisfying $g_{\phi
} (\bmactual\mid\byactual) > 0$, $L_{3, \phi} (\btheta)$ is
proportional to $L_2 (\btheta)$; and \textup{(iii)} if $\exists\bphi\in
\Omega_{\phi}$ such that $g_{\phi} (\bmactual\mid\byactual) > 0$,
then $L_4 (\btheta)$ is a special case of $L_{3, \phi} (\btheta)$
and, hence, $L_4 (\btheta)$ is proportional to $L_2 (\btheta)$.
\end{theorem}

\begin{pf}
As $\Omega_{\theta, \phi} = \Omega_{\theta} \times\Omega_{\phi
}$, it follows that whenever $\bphi\in\Omega_{\phi}$ and $\btheta
\in\Omega_{\theta}$, then $(\btheta, \bphi) \in\Omega_{\theta,
\phi}$, and so $\delta\{(\btheta,\bphi),\Omega_{\theta,\phi}\}= 1$.
So, for $\bphi\in\Omega_{\phi}$ and $\btheta\in\Omega_{\theta}$,
%
\begin{eqnarray}
\label{eqjointdensity}
L_1 (\btheta, \phi) & = & \int f_{\theta} (\mathbf y )
g_{\phi} (\bmactual\mid\mathbf y ) r(\mathbf y, \byactual, \bmactual) \,d \mathbf
y
\\
\label{eqneedsMAR1}
& = & g_{\phi} (\bmactual\mid\byactual) \int f_{\theta} (\mathbf y )
r(\mathbf y, \byactual, \bmactual) \,d \mathbf y
\\
\label{eqfactorises}
& = & L_5 (\bphi) L_2 (\btheta),
\end{eqnarray}
where
\[
L_5 (\bphi) = g_{\phi} (\bmactual\mid\byactual)
\]
is a function of $\bphi$ only.
Hence, (i) is true.
Note that line (\ref{eqneedsMAR1}) follows because of realised MAR.

If realised MAR holds and $\Omega_{\theta, \phi} = \Omega_{\theta}
\times\Omega_{\phi}$, line (\ref{eqneedsMAR1}) is equal to $L_{3, \phi}
(\btheta)$. Since $g_{\phi} (\bmactual\mid\byactual)$ is not a function
of~$\btheta$, it then follows that $L_{3, \phi} (\btheta)$ is
proportional to $L_2 (\btheta)$ for any $\bphi\in\Omega_{\phi}$ such
that $g_{\phi} (\bmactual\mid\byactual) > 0$. So, (ii) is true.

Likewise, when the data are realised MAR and $\Omega_{\theta, \phi}
= \Omega_{\theta} \times\Omega_{\phi}$,
\[
L_4 (\btheta) = \int f_{\theta} (\mathbf y ) r(\mathbf y,
\byactual, \bmactual) \,d \mathbf y \times\max_{\bphi\in\Omega_{\phi}}
g_{\phi}
(\bmactual\mid\byactual).
\]
The function $\max_{\bphi\in\Omega_{\phi}}
g_{\phi} (\bmactual\mid\byactual)$ does not depend on~$\btheta$.
Moreover, it is nonzero when $\exists\bphi\in\Omega_{\phi}$ such
that $g_{\phi} (\bmactual\mid\byactual) > 0$.
So, $L_4 (\btheta) = L_{3, \hat{\phi}} (\btheta)$, where $\hat
{\bphi}$ is the value of $\bphi$ that maximises $g_{\phi} (\bmactual
\mid\byactual)$.
Hence, (iii) is true.
\end{pf}

In the literature, this factorisation of the joint likelihood and this
proportionality of likelihoods have been used as a basis for defining
when the missingness mechanism can be ignored when performing
direct-likelihood inference.
Rubin~\cite{Rubin1976}, for example, used the proportionality of likelihoods to
write: ``When making direct-likelihood or Bayesian inferences about
$\btheta$, it is appropriate to ignore the process that causes missing
data if the missing data are missing at random and the parameter of the
missing data process is ``distinct'' from~$\btheta$''.
Anscombe~\cite{Anscombe1964}
wrote that when the joint likelihood for a parameter of interest
$\btheta$ and a nuisance parameter $\bphi$ factorises into two
components, such that each parameter appears in only one component,
information on each factor can be considered separately.
The same was written by Hinde and Aitkin~\cite{Hinde1987}.
Royall~\cite{Royall1997}
called the component involving $\btheta$ the ``likelihood for
$\btheta$'' and said that the relative support for any two values of
$\btheta$ is given by the ratio of the values of this likelihood
evaluated at those two $\btheta$ values.
Edwards~\cite{Edwards1970} supported the use of the profile
likelihood when the joint likelihood factorises.
He wrote: ``since the value of $\bphi$ is irrelevant to our inference
on $\btheta$, replacing $\bphi$ in [the joint likelihood] by its
maximum likelihood estimate will not invalidate the likelihood''.
Kalbfleisch and Sprott~\cite{Kalbfleisch1970} agreed with Edwards.
When comparing inference for $\btheta$ using $L_1$ and $L_2$ in
situations where the two may give different answers, Heitjan
\cite{Heitjan1991,Heitjan1993}, pages 1103 and 2249,
interpreted inference for $\btheta$ using $L_1$ as meaning inference
using the profile likelihood.
Tsou and Royall~\cite{Tsou1995}
considered the strength of evidence in the presence of a nuisance
parameter as being the strength of evidence that would be in the data
if the value of the nuisance parameter were known.
That is, they considered the strength of evidence to be the particular
fixed-$\bphi$ likelihood for $\btheta$ corresponding to the true
value of $\bphi$.

All these authors, therefore, provide justification for interpreting
Theorem~\ref{theo1} as meaning that direct-likelihood inference about $\btheta$
can be performed using $L_2$ when the data are realised MAR and
$\btheta$ and $\bphi$ are distinct parameters.

To picture the effect of realised MAR and distinctness of parameters on
the joint likelihood $L_1(\btheta, \bphi)$, it is helpful to consider
a joint model where $\btheta$ and $\bphi$ are\vadjust{\goodbreak} both scalar parameters.
The graph of $L_1$ is then a surface in three dimensions lying above a
$(\theta,\phi)$ plane.
The realised MAR condition imposes geometric structure on this surface
[evident from equations (\ref{eqneedsMAR1}) and (\ref
{eqfactorises})] such that curves obtained from the surface by fixing
$\phi$ at various values are all proportional, simply being copies of
$L_2$ scaled by the conditional probability of realising the observed
missingness pattern under the given $\bphi$ value.
The function $L_1$ is, however, only defined for values of $(\btheta,
\bphi)$ in $\Omega_{\theta,\phi}$.
Hence, the curve formed from the $L_1$ by fixing $\bphi$ may be
undefined for some values of $\btheta$ where the $L_2$ curve is defined.
So, one can think of each curve formed from $L_1$ by fixing $\bphi$ as
being a proportional copy of $L_2$ with, potentially, one or more
sections omitted.
The assumption of distinct parameters ensures that such ``omitted''
sections do not exist, and therefore that the curves are proportional
at all $\btheta$ values in $\Omega_{\theta}$.

So far, we have considered the elimination of $\bphi$ as a nuisance
parameter. As discussed in Section~\ref{sectinferencetypes}, when a
likelihood interval is required for a single element,~$\theta_1$, of a
vector parameter, $\btheta$, the other parameters, $\btheta_2$, are
also nuisance parameters and must be eliminated. If $\btheta_2$ is
eliminated from $L_2 (\btheta)$ and $L_4 (\btheta )$ using the profile
likelihood method, the proportionality of $L_4 (\btheta)$ and $L_2
(\btheta)$ also ensures the proportionality of the profile likelihoods
for $\theta_1$ derived from $L_2 (\btheta)$ and $L_4 (\btheta)$. Hence,
the likelihood intervals for $\theta_1$ obtained from $L_2$ and $L_4$
will be the same. We discuss the use of conditional likelihood in
Section~\ref{sectconditional}.

\subsection{Bayesian Inference}

Consider Bayesian inference accounting for the missingness mechanism.
Let $p_{\theta, \phi} (\btheta, \bphi)$ denote the joint prior
distribution of $(\btheta, \bphi)$ and let $p_{\theta} (\btheta)$
denote the corresponding marginal prior distribution of $\btheta$.
The missingness mechanism is said to be ignorable for Bayesian
inference if the marginal posterior distribution of $\btheta$ obtained
from modelling both the complete data, $\bY$, and the missingness
pattern, $\bM$, is equal to the posterior for $\btheta$ obtained by
modelling $\bY$ alone.
The main work in this area can be summed up by the following theorem.

\begin{theorem}\label{theo2}
Suppose that (1) the data are realised MAR and (2) $\btheta$ and
$\bphi$ are a priori independent.
The posterior distribution of $\btheta$ that results from using the
likelihood $L_2 (\btheta)$ and the prior $p(\btheta)$ is the same as
the posterior distribution that results from using likelihood $L_1
(\btheta, \bphi)$ and prior $p_{\theta, \phi} (\btheta, \bphi)$.
\end{theorem}

\begin{pf}
When $L_1 (\btheta, \bphi)$ and $p_{\theta, \phi}(\btheta, \bphi
)$ are used, the posterior distribution of $(\btheta, \bphi)$ is
proportional to $p_{\theta, \phi}(\btheta, \bphi) L_1 (\btheta,
\bphi)$.\vadjust{\goodbreak}
If $\btheta$ and $\bphi$ are a priori independent, $p_{\theta, \phi
} (\btheta, \bphi)$ factorises as $p_{\theta} (\btheta) p_{\phi}
(\bphi)$, where $p_{\phi} (\bphi)$ is the marginal prior for $\bphi$.
If, furthermore, the data are realised MAR, it follows from
equation (\ref{eqfactorises}) that the posterior distribution of
$(\btheta, \bphi)$ is proportional to $p_{\phi} (\bphi) L_5(\bphi)
p_{\theta} (\btheta) L_2 (\btheta)$.
Since $p_{\phi} (\bphi) L_5(\bphi)$ is a function of $\bphi$ only,
the marginal posterior distribution of $\btheta$ is proportional to
$p_{\theta} (\btheta) L_2 (\btheta)$.
This is the same posterior distribution that is obtained if $L_2
(\btheta)$ and $p_{\theta} (\btheta)$ are used.
\end{pf}

\subsection{General Frequentist Inference}

From the joint model, for any $\bphi\in\Omega_{\phi}$ for which
$\exists\mathbf y $ such that $f_{\theta} (\mathbf y ) g_{\phi} (\bmactual
\mid\mathbf y ) > 0$, the conditional distribution of $o(\bY, \bM)$
given $\bM= \bmactual$ is
%
\begin{eqnarray}\label{eqconditionaldist}
&&\int f_{\theta} (\bu) g_{\phi} (\bmactual\mid\bu) r(\bu, \mathbf
y, \bmactual) \,d \bu\nonumber\\[-8pt]\\[-8pt]
&&\quad\hspace*{0pt}{}\Big/\int f_{\theta} (\bu) g_{\phi} (\bmactual\mid
\bu) \,d \bu.\nonumber
\end{eqnarray}
In general, this distribution may depend on $\bphi$. Let $t\{ o(\bY,
\bM), \bM\}$ be a function of $o(\bY, \bM)$ and $\bM$. Rubin~\cite{Rubin1976}
called the distribution of $t\{ o(\bY, \bM), \bM \}$ given $\bM=
\bmactual$ implied by the distribution of $o(\bY,\bM)$ given $\bM=
\bmactual$ in expression (\ref {eqconditionaldist}) the ``correct
conditional sampling distribution'' of $t\{ o(\bY, \bM), \bM\}$. In
general, the distribution given by (\ref{eqconditionaldist}) is not
equal to
%
\begin{equation}\label{eqmarginaldist}
\int f_{\theta} (\bu) r(\bu, \mathbf y, \bmactual) \,d \bu
\end{equation}
and so, in general, the conditional distribution of $o(\bY, \bM)$
given $\bM= \bmactual$ is not that given by expression (\ref
{eqmarginaldist}).
Nevertheless, the latter distribution is the distribution that
corresponds to likelihood $L_2 (\btheta)$.
Heitjan and Basu~\cite{Heitjan1996} called the distribution of\break
$t\{ o(\bY, \bM), \bM\}$ given $\bM= \bmactual$ implied by
the distribution in (\ref{eqmarginaldist}) the ``potentially incorrect
sampling distribution'' of $t\{ o(\bY, \bM), \bM\}$.

\begin{theorem}\label{theo3}
When the data are realised MCAR and $\exists\mathbf y $ such that
$f_{\theta} (\mathbf y ) g_{\phi} (\bmactual\mid\mathbf y ) > 0$, the
potentially incorrect sampling distribution is equal to the correct
conditional sampling distribution.
\end{theorem}

\begin{pf}
If the data are realised MCAR, then for each value of $\bphi$ the
value of $g_{\phi} (\bmactual\mid\mathbf y )$ does not depend on $\mathbf y $.
Hence, expression (\ref{eqconditionaldist}) reduces to
expression~(\ref{eqmarginaldist}).
\end{pf}

Note that in Theorem~\ref{theo3} repeated sampling is conditional on the realised
missingness pattern, that is, conditional on $\bM= \bmactual$.
Little~\cite{Little1976} argued that it is wrong to condition on $\bM= \bmactual
$, as $\bM$ is not an ancillary statistic for $\btheta$ unless the
stronger condition of everywhere\vadjust{\goodbreak} MCAR is satisfied.
Rubin~\cite{Rubin1976b} disagreed, saying that ``the usual
definition of ancillary (Cox and Hinkley~\cite{Cox1974}, page 35)
is incorrect for inference about $\btheta$ and should be modified to
be conditional on the observed value of the ancillary statistic''.
Heitjan~\cite{Heitjan1997} continued this discussion,
introducing the concept of an observed ancillary statistic and agreeing
with Rubin's assertion that the missingness pattern can be conditioned
upon when the data are realised MCAR.
As Rubin noted, although Theorem~\ref{theo3} might be regarded as a statement
about when the missingness mechanism can be ignored, the realised
missingness pattern itself is not ignored, because the repeated
sampling is conditional on it.

As mentioned in Section~\ref{sectinferencetypes}, repeated sampling
may be conditional on a function of $\bY$.
We discuss this in Section~\ref{sectconditional}.

\subsection{Frequentist Likelihood Inference}
\label{sectfrequentistlikelihood}

Since frequentist likelihood inference is a special case of general
frequentist inference, Theorem~\ref{theo3} still applies. However, for
frequentist likelihood inference a further result can be obtained when
the data are everywhere MAR and $\btheta$ and $\bphi$ are distinct.
When the data are not realised MCAR, $\bM$ is not observed ancillary,
and so repeated sampling should not be conditional on $\bM$. However,
if the data are everywhere MAR and $\Omega_{\btheta, \bphi} =
\Omega_{\btheta} \times\Omega_{\bphi}$, it follows from
Theorem~\ref{theo1} that $L_2 (\btheta)$, $L_{3, \phi} (\btheta)$ and
$L_4 (\btheta)$ are proportional not only in the realised sample but
also in repeated samples. Therefore, the MLE of $\btheta$, the
estimated variance of this MLE calculated from the observed information
matrix, likelihood intervals for~$\btheta$, and likelihood-ratio, Wald
and score test statistics for hypotheses concerning $\btheta$ will be
the same in both the realised and repeated samples whether calculated
using $L_2$ or $L_1$. That is, the same frequentist likelihood
inference for $\btheta$ will be made whether one uses $L_2$ or $L_1$.

A similar result applies for Bayesian point estimators and credible
intervals. Suppose that the data are everywhere MAR and, for every
possible data vector $\bY$ and missingness pattern $\bM$, the prior for
$(\btheta, \bphi)$ in the joint model can be written as $p(\btheta,
\bphi) = p(\btheta) \times p(\bphi)$, where $p(\btheta)$ is the prior
for $\btheta$ in the model that ignores the missingness pattern. Then,
for every possible $(\bY, \bM)$, the posterior distribution for
$\btheta$ derived from the likelihood $L_1$ and prior $p(\btheta,
\bphi)$ of the joint model is the same as that derived from the
likelihood $L_2$ and prior $p(\btheta)$ of the model that ignores the
missingness pattern. Consequently, under these assumptions, the
repeated-sampling properties of Bayesian point estimators and credible
intervals for $\btheta$ in repeated samples will be the same whether
one uses $L_1$ and $p(\btheta, \bphi)$ and integrates over $\bphi$ or
one uses $L_2$ and $p(\btheta)$.


One important caveat needs to be stated.
Standard errors can, in general, be calculated using either the
expected or the observed information.
When the data are everywhere MAR and $\btheta$ and $\bphi$ are
distinct, the expected information from $L_2$ should not be used
naively~\cite{Kenward1998}.
Using this expected information is only appropriate under the stronger
assumption that the data are everywhere MCAR.
It is recommended that the observed information be used instead
\cite{Kenward1998}.

\section{Conditional Likelihood and Repeated Sampling}
\label{sectconditional}

We now consider (1) conditional likelihoods and (2) repeated sampling
conditional not only on $\bM= \bmactual$ but also on some function of
$\bY$.
Let $\bX= b(\bY)$ denote the function of $\bY$ being conditioned on
and $\tilde{\mathbf x }$ denote the realised value of $\bX$.

First, consider the use of conditional likelihood.
One example of the use of a conditional likelihood is where data $\bY$
consist of a set of covariates and an outcome for a sample of
individuals and this outcome is regressed on the covariates.
When the covariates are fully observed, there is no need to specify a
likelihood for all of $\bY$; instead, a likelihood for the outcomes
conditional on the covariates is sufficient.
Here, $\bX$ consists of the covariates.
A~second example is conditional logistic regression for matched
case-control data, where the likelihood is conditional on the number of
cases and controls in each matched set.
So here, $\bX$ consists of these numbers of cases and controls.

Assume that either $\tilde{\mathbf x }$ is observed or $\int f_{\theta}
(\mathbf y \mid\mathbf x = \tilde{\mathbf x }) r(\mathbf y, \byactual,
\bmactual) \,d \mathbf y $ does not
depend on the value of the missing part of $\tilde{\mathbf x }$.
In equations (\ref{eqL1})--(\ref{eqL4}), $f_{\theta} (\mathbf y )$
should be replaced by $f_{\theta} (\mathbf y \mid\mathbf x = \tilde{\mathbf x
})$, the conditional probability distribution of $\bY$ given $\bX=
\tilde{\mathbf x }$.
Theorem~\ref{theo1} then still applies.
Moreover, if the data are everywhere MAR, then $L_2 (\btheta)$ and
$L_4 (\btheta)$ [both with $f_{\theta} (\mathbf y )$ replaced by
$f_{\theta} (\mathbf y \mid\mathbf x = \tilde{\mathbf x })$] will be
proportional not only in the realised sample but also in repeated samples.
Note that this repeated sampling is conditional on $\bX= \tilde{\mathbf x
}$ but not on $\bM= \bmactual$.

Second, consider repeated sampling conditional on $\bX= \tilde{\mathbf x
}$ and $\bM= \bmactual$.
Assume that either $\tilde{\mathbf x }$ is observed or the distribution of
$t\{ o(\bY, \bmactual), \bmactual\}$, given $\bM= \bmactual
$ and $\bX= \tilde{\mathbf x }$ implied by the distribution $\int
f_{\theta} (\mathbf y \mid\mathbf x = \tilde{\mathbf x }) r(\mathbf y,
\byactual, \bmactual) \,d \mathbf y $,
does not depend on the value of the missing part of $\tilde{\mathbf x }$.
In equations (\ref{eqL1})--(\ref{eqL4})\vadjust{\goodbreak} and (\ref
{eqmarginaldist}), $f_{\theta} (\mathbf y )$ should be replaced by
$f_{\theta} (\mathbf y \mid\mathbf x = \tilde{\mathbf x })$, and $f_{\theta}
(\bu)$ in equation (\ref{eqconditionaldist}) should be replaced by
$f_{\theta} (\bu\mid\mathbf x = \tilde{\mathbf x })$.
Theorem~\ref{theo3} then continues to apply if ``$f_{\btheta} (\mathbf y ) g_{\phi}
(\bmactual\mid\mathbf y ) > 0$'' is replaced by ``$f_{\theta} (\mathbf y
\mid\mathbf x = \tilde{\mathbf x }) g_{\phi} (\bmactual\mid\mathbf y ) > 0$
and $b(\mathbf y ) = \tilde{\mathbf x }$''.
Moreover, the realised MCAR condition in Theorem~\ref{theo3} can be replaced by
the following weaker condition: $\forall\bphi$, $g_{\phi} (\bmactual
\mid\mathbf y ) = g_{\phi} (\bmactual\mid\mathbf y ^*)$ $\forall\mathbf y,
\mathbf y ^*$ such that $b(\mathbf y ) = b(\mathbf y ^*) = \tilde{\mathbf x }$.
In the special case of repeated-measures data with fully observed
covariates and $\bX$ being these covariates, the everywhere version of
this weaker condition has been called\break ``covariate-dependent MCAR''
\cite{Little1995,Wood2004}.

\section{Discussion}
\label{sectdiscussion}

In this article we have highlighted inconsistencies in the use of the
terms ``missing at random'' and ``likelihood inference'', and clarified the
conditions required for ignorability of the missingness mechanism.
We urge those writing about missing data to be clearer with respect to
the assumptions being used and to employ clear terminology when
describing approaches to inference, in particular, to make the
distinction between direct-likelihood and frequentist likelihood concepts.

Rubin~\cite{Rubin1976} used the term ``ignorable'' to mean that
two likelihoods, one derived from a model for the data alone
and one derived from a joint model for the data and the missingness
pattern, are proportional or that two sampling distributions, the
``potentially incorrect'' and correct conditional distributions, are equal.
In Section~\ref{sectignorability} we explained how this implies that
certain inferences for $\btheta$ from the two models are the same.
In this interpretation, ignorability is a property of the \textit{assumed}
missingness model.
Whether this assumed model is correctly specified is not relevant.
This interpretation of ``ignorability'' may not be universal, however.
As we saw in Section~\ref{sectreview}, some writers have omitted the
parameter $\bphi$ from their definition of MAR.
Rather than refer to a model for the missingness mechanism, they appear
to have been referring to the ``true'' missingness mechanism (which is
usually unknown).
Such writers may have interpreted ignorability to mean that using $L_2
(\btheta)$ for frequentist likelihood (or frequentist Bayesian)
inference will be valid, that is, will yield consistent MLEs (or
posterior modes), consistent variance estimators, confidence (or
credible) intervals with asymptotic nominal coverage, etc.
Theorem~\ref{theo1} implies the following result.
Suppose that the ``true'' missingness mechanism is $P(\bM= \bem\mid\bY
= \mathbf y )$ and that $P(\bM= \bem\mid\bY= \mathbf y ) = P(\bM= \bem
\mid\bY= \mathbf y ^*)$\vadjust{\goodbreak} $\forall\bem, \mathbf y, \mathbf y ^*$ such that $o
(\mathbf y, \bem) = o(\mathbf y ^*, \bem)$.
A~hypothetical analyst who knew this ``true'' missingness mechanism and
wanted to make inference for $\btheta$ taking missingness into account
would use the likelihood $\int f_{\theta} (\mathbf y ) P(\bM= \bmactual
\mid\bY= \mathbf y ) r(\mathbf y, \byactual, \bmactual) \,d \mathbf y $
and, by so doing, obtain valid
frequentist likelihood (or frequentist Bayesian) inference.
Theorem~\ref{theo1} implies that $L_2 (\btheta)$ is proportional to this
likelihood, and hence that valid frequentist likelihood (or frequentist
Bayesian) inference would also be obtained using~$L_2$.

Despite MAR plus distinctness of parameters being presented in Little
and Rubin~\cite{Little2002} as the definition of ignorability (Definition 6.4),
Theorems~\ref{theo1} and~\ref{theo2} only give \textit{sufficient} conditions for when it is
appropriate to ignore the missingness mechanism when making
direct-likelihood and Bayesian inferences, respectively.
In the case of direct-likelihood inference, Theorem~\ref{theo1} is concerned with
sufficient conditions for $L_{3, \phi} (\btheta)$, the fixed-$\bphi$
likelihood for $\btheta$, to be proportional to $L_2 (\btheta)$, the
likelihood for $\btheta$ ignoring the missing data mechanism.
It is conceivable that, even in the absence of realised MAR, there may
be a restricted set of $\bphi$ values for which $L_2(\btheta)$ is
proportional to $L_{3, \phi}(\btheta)$, and for this restricted set
to contain the ``true'' $\bphi$ value.
If so, it would be appropriate to ignore the missingness mechanism even
though realised MAR does not hold.
Lu and Copas~\cite{Lu2004} showed that, when $\btheta$ and $\bphi$ are
distinct \textit{and} the family of distributions $f_\theta(\mathbf y )$ form
a complete class, everywhere MAR is necessary and sufficient for
ignorability in frequentist likelihood inference.
It is straightforward to adapt their proof to show that when $\btheta$
and $\bphi$ are distinct and the family of distributions
$f_\theta( \mathbf y \mid o(\mathbf y, \bem) = o(\byactual,
\bmactual) )$
form a complete class, then realised MAR is necessary and sufficient
for ignorability in direct likelihood inference (we include a proof in
the \hyperref[app]{Appendix}).
Furthermore, there may conceivably be other ways, apart from that of
using a fixed-$\bphi$ likelihood, to extract a likelihood for $\btheta
$ from $L_1$, ways which may not require realised MAR and parameter
distinctness in order for the extracted likelihood to be proportional
to $L_2$.
In the case of Theorem~\ref{theo2}, it is conceivable that independence of the
posterior distributions for $\btheta$ and $\bphi$ may be a stronger
condition than is necessary, and it seems to still be an open question
whether there are substantially weaker conditions under which it is
appropriate to ignore the missingness mechanism when performing
Bayesian inference.

Note that the concept of missing data has been generalised to that of
``coarsened'' data~\cite{Heitjan1993}.
When data are coarsened, data values\vadjust{\goodbreak} are not necessarily either
observed or missing, instead one observes a set of values that is known
to contain the realised values.
Censored survival data are an example of coarsened data: a survival
time may be known to be greater than a given (censoring) time but not
known exactly.

We conclude with some brief remarks on the potential practical
implications of this work.
Our review of the literature on the theory of missing data methods has
highlighted a number of inconsistencies and a lack of clarity with
respect to key definitions such as MAR and ignorability.
We believe that these issues have clouded the development and broader
understanding of methods in this area, partly because they intersect in
considerable measure with issues in the foundations of statistical inference.
Although the original definition of MAR (our ``realised MAR'') provides a
clear basis for thinking about direct likelihood and Bayesian
inferences, the majority of statistical practice is concerned with
frequentist evaluations.
Even those who emphasise the Bayesian interpretation of particular
analyses are generally interested in repeated-sampling performance of
procedures.
Incomplete data methods that do not explicitly model the missing data
mechanism (i.e., that assume ignorability) cannot be guaranteed to
perform validly in repeated samples except under an ``everywhere'' MAR
assumption.
The restrictiveness of this assumption does not seem to be well
understood, especially in complex problems with nonmonotone patterns of
missingness~\cite{Robins1997,Potthoff2006}.
More importantly, further work is needed on methods to more effectively
and systematically characterise the potential sensitivity of inferences
to departures from the MAR assumption.
Meanwhile, users of missing data methods need to be reminded that
methods that assume ignorability provide tractable analyses only at the
cost of untestable assumptions.

It is also important to consider that when there are missing data,
there is more than one possible target of inference.
Diggle et al.~\cite{Diggle2007} discuss alternative possible
study objectives and targets of inference that are relevant to those objectives.

Much recent research in methods for handling missing data has
considered issues that are specific to the structure of the problem.
For example, missingness in outcomes poses different challenges than
does missingness in covariate values, and longitudinal (repeated
measures) data present specific issues of their own.
We believe that it should be possible to tackle these problems with
greater clarity if the fundamental assumptions\vadjust{\goodbreak} about missing data
mechanisms and their connection with the concept of ignorability are
better understood.

\begin{appendix}\label{app}
\section*{Appendix}

Here we show that when $\btheta$ and $\bphi$ are distinct and the
family of distributions
$f_\theta( \mathbf y \mid o(\mathbf y, \bem) = o(\byactual,
\bmactual) )$
form a complete class, then realised MAR is necessary and sufficient
for ignorability.

Let $\mis( \bY, \bM)$ denote the subvector of $\bY$ consisting of
the elements whose corresponding elements of $\bM$ equal zero.
So, $\mis(\bY, \bM)$ contains the missing elements of $\bY$.
For any fixed value $\bem$ of $\bM$, $f_{\theta} (\mathbf y )$ can be
written as
%
\renewcommand{\theequation}{\arabic{equation}}
\begin{equation}\label{eqfactorisewithm}
\qquad
f_{\theta} (\mathbf y ) = f_{1, \theta} \bigl\{ o(\mathbf y, \bem) \bigr\}
f_{2,
\theta} \bigl\{ \mis(\mathbf y, \bem) \mid o(\mathbf y, \bem) \bigr\}.
\end{equation}
Thus, choosing $\bem= \bmactual$ in equation (\ref
{eqfactorisewithm}), $L_1$ can be written as
\begin{eqnarray*}
L_1 (\btheta, \bphi) & = & \int f_{1, \theta} \bigl\{ o(\mathbf y,
\bmactual) \bigr\} f_{2, \theta} \bigl\{ \mis(\mathbf y, \bmactual) \mid
o(\mathbf y,
\bmactual) \bigr\} \\
&&\hspace*{9.5pt}{}\cdot g_{\phi} (\bmactual\mid\mathbf y ) r(\mathbf y,
\byactual, \bmactual) \,d \mathbf y
\\
& = & f_{1, \theta} \bigl\{ o(\byactual, \bmactual) \bigr\} \int
f_{2, \theta} \bigl\{ \mis(\mathbf y, \bmactual) \mid o(\byactual,
\bmactual)
\bigr\}\\
&&\hspace*{72.5pt}{}\cdot
g_{\phi} (\bmactual\mid\mathbf y ) r(\mathbf y, \byactual,
\bmactual
) \,d \mathbf y
\end{eqnarray*}
%
%
and $L_2$ can be written as $L_2 (\btheta) = f_{1, \theta} \{ o
(\byactual, \bmactual) \}$.

\begin{theorem*}
Suppose that $\Omega_{\theta, \phi} = \Omega_{\theta} \times
\Omega_{\phi}$, that $f_{2, \theta} \{ \mis(\mathbf y, \bmactual)
\mid o(\byactual, \bmactual) \}$ is complete, and that $g_{\phi}
(\bmactual\mid\byactual) > 0$ for all $\bphi\in\Omega_{\phi}$.
Then $L_1 (\btheta, \bphi)$ is proportional to $L_2 (\btheta)$ for
any $\bphi\in\Omega_{\phi}$ if and only if realised MAR holds.
\end{theorem*}
\begin{pf}
The ``if'' argument holds because of Theorem~\ref{theo1}.
So, consider the ``only if'' argument.
Suppose that $L_1 (\btheta, \bphi)$ is proportional to $L_2 (\btheta
)$ for any $\bphi\in\Omega_{\phi}$.
Then it must be true that for all $\bphi\in\Omega_{\phi}$,
%
\begin{equation}\label{eqQdefinition}
\int f_{2, \theta} \bigl\{ \mis(\mathbf y, \bmactual) \mid o(\byactual,
\bmactual) \bigr\} g_{\phi} (\bmactual\mid\mathbf y ) r(\mathbf y,
\byactual,
\bmactual) \,d \mathbf y\hspace*{-25pt}
\end{equation}
cannot depend on $\btheta$.
Hence, we can denote expression (\ref{eqQdefinition}) as $Q \{
\bmactual, o(\byactual, \bmactual), \bphi\}$.

By definition,
\begin{eqnarray*}
&&\int f_{2, \theta} \bigl\{ \mis(\mathbf y, \bmactual) \mid o(\byactual,
\bmactual) \bigr\} g_{\phi} (\bmactual\mid\mathbf y ) r(\mathbf y,
\byactual,
\bmactual) \,d \mathbf y \\
&&\quad{}- Q \bigl\{\bmactual, o(\byactual, \bmactual), \bphi
\bigr\} =
0.
\end{eqnarray*}
%
%
So,
\begin{eqnarray*}
&& \int f_{2, \theta} \bigl\{ \mis(\mathbf y, \bmactual) \mid o (\byactual,
\bmactual) \bigr\} g_{\phi} (\bmactual\mid\mathbf y ) r(\mathbf y,
\byactual,
\bmactual) \,d \mathbf y
\\
&&\quad{} - Q \bigl\{\bmactual, o(\byactual, \bmactual), \bphi\bigr\} \\
&&\hspace*{22.2pt}{}\cdot \int
f_{2,
\theta} \bigl\{ \mis(\mathbf y, \bmactual) \mid o(\byactual, \bmactual)
\bigr\} r(\mathbf y, \byactual, \bmactual) \,d \mathbf y = 0
\end{eqnarray*}
for all $\bphi\in\Omega_{\phi}$.
It then follows that
\begin{eqnarray*}
&&
\int f_{2, \theta} \bigl\{ \mis(\mathbf y, \bmactual) \mid o(\byactual,
\bmactual) \bigr\} \\
&&\hspace*{9pt}{}\cdot \bigl[ g_{\phi} (\bmactual\mid\mathbf y ) - Q \bigl\{
\bmactual, o(\byactual, \bmactual), \bphi\bigr\} \bigr] r(\mathbf y,
\byactual,
\bmactual) \,d \mathbf y = 0
\end{eqnarray*}
%
%
for all $\bphi\in\Omega_{\phi}$.
So, if $f_{2, \theta} \{ \mis(\mathbf y, \bmactual) \mid o
(\byactual, \bmactual) \}$ is complete, then
$Q \{\bmactual, o(\byactual, \bmactual), \bphi\} = g_{\phi}
(\bmactual\mid\mathbf y )$ for all $\bphi\in\Omega$ and for all $\mathbf y
$ such that $o(\mathbf y, \bmactual) = o(\byactual, \bmactual)$.
Therefore, $g_{\phi} (\bmactual\mid\mathbf y )$ cannot depend on $\mis
(\byactual, \bmactual)$,
that is, the data are realised MAR.
\end{pf}
\end{appendix}

\section*{Acknowledgements}

We thank Professor Mike Kenward for very useful discussions and for
comments on a draft manuscript, and anonymous reviewers for their
helpful suggestions.

S. R. Seaman is funded by MRC Grants U1052 60558 and
MC\_US\_A030\_0015. D. Jackson is funded by U1052 60558. J. B. Carlin
and J. C. Galati wish to acknowledge the support of research Grant
607400 from the Australian National Health and Medical Research
Council, and support provided to the MCRI from the Victorian
Government's Operational Infrastructure Support Program.



\end{document}